\documentclass[prb,twocolumn,preprintnumbers,amsmath,amssymb]{revtex4}
\usepackage{graphicx}
\usepackage{dcolumn}
\usepackage{bm}

\begin{document}


\title{Fe and N self-diffusion in amorphous FeN:\\ A SIMS and neutron reflectivity study}

\author{S. Chakravarty$^{1,a}$}\author{M. Gupta$^{2}$}\email{mgupta@csr.ernet.in}\author{A. Gupta$^{1}$, S.
Rajagopalan,$^{3}$ A. K. Balamurugan,$^{3}$ A. K. Tyagi,$^{3}$
U.P. Deshpande,$^{1}$ M. Horisberger$^{4}$ and T.
Gutberlet$^{4,b}$}

\affiliation{ $^{1}$UGC-DAE Consortium for
Scientific Research, Khandwa Road, Indore 452 017,
India\\    $^{2}$UGC-DAE Consortium for Scientific Research, R-5 Shed, BARC, Mumbai 400 085, India\\
$^{3}$Surface Science Section, MSD, IGCAR, Kalpakkam 603 102,
India\\   $^{4}$Laboratory for Neutron Scattering, ETH Z\"{u}rich
and Paul Scherrer Institute, CH-5232 Villigen PSI,
Switzerland\\$^{a}$Present address: Clausthal University of
Technology, D-38678 Clausthal-Zellerfeld, Germany\\$^{b}$Present
address:Forschungszentrum J\"{u}lich GmbH, J\"{u}lich Centre for
Neutron Science at FRM II, 85747 Garching, Germany}
\date{\today}

\begin{abstract}
Simultaneous measurement of self-diffusion of iron and nitrogen in
amorphous iron nitride (Fe$_{86}$N$_{14}$) using secondary ion
mass spectroscopy (SIMS) technique has been done. In addition
neutron reflectivity (NR) technique was employed to study the Fe
diffusion in the same compound. The broadening of a tracer layer
of $^{57}$Fe$_{86}$$^{15}$N$_{14}$ sandwiched between
Fe$_{86}$N$_{14}$ layers was observed after isothermal vacuum
annealing of the films at different temperatures in SIMS
measurements. And a decay of the Bragg peak intensity after
isothermal annealing was observed in
[Fe$_{86}$N$_{14}$/$^{57}$Fe$_{86}$N$_{14}$]$_{10}$ multilayers in
NR. Strong structural relaxation of diffusion coefficient was
observed below the crystallization temperature of the amorphous
phase in both measurements. It was observed from the SIMS
measurements that Fe diffusion was about 2 orders of magnitude
smaller compared to nitrogen at a given temperature. The NR
measurements reveal that the mechanism of Fe self-diffusion is
very similar to that in metal-metal type metallic glasses. The
structural relaxation time for Fe and N diffusion was found
comparable indicating that the obtained relaxation time
essentially pertain to the structural relaxation of the amorphous
phase.

\end{abstract}

\maketitle

\section{\label{sec:level1}Introduction}
FeN systems show a variety of structures with composition.
Technologically, FeN is an interesting system with applications in
many industries.\cite{Wang:JMMM04,Wei:NIMB91,Wit:PRL94} Recently
amorphous phases of iron nitride were prepared using reactive
sputtering with nitrogen as reactive
gas.\cite{Gupta:amorph_PRB05,Ranu:PRB:2006,gupta:214204:PRB02}
Amorphous phases of FeN were obtained in nitrogen poor as well as
nitrogen rich
phases.~\cite{Gupta:amorph_PRB05,Ranu:PRB:2006,gupta:214204:PRB02}
The nitrogen poor amorphous phases produced using this method were
not stable beyond a temperature of 523 K and undergo
crystallization when heated above this temperature. Within the
amorphous phase conventional amorphous alloys are known to undergo
structural relaxation.

In an earlier work, iron diffusion in nitrogen rich amorphous FeN
alloy produced using ion beam sputtering technique was studied and
it was found that Fe diffusion demonstrates strong relaxation
effects similar to a metal-metalloid
system.~\cite{gupta:214204:PRB02} Structurally, the nitrogen rich
and the nitrogen poor FeN phases are different though in both the
cases they are amorphous. The broad diffuse maxima in case of
nitrogen poor amorphous FeN appears around the same position as in
case of bcc-Fe (2\,$\theta$ $\sim$ 44\,$^{\circ}$, with
1.54\,{\AA} x-rays), whereas in case of nitrogen rich amorphous
FeN the position of this diffuse maximum appears at much lower
angle (2\,$\theta$ $\sim$ 37\,$^{\circ}$, with 1.54\,{\AA}
x-rays). This indicates that structurally nitrogen rich and
nitrogen poor phases are very different. Further in case of
nitrogen poor FeN phases nitrogen predominately occupies
interstitial positions and in case of nitrogen rich FeN phases
covalent bonds between Fe and N are formed. In order to study the
phenomenon of structural relaxation and self-diffusion in the
system it is desirable to study both Fe as well as N
self-diffusion.

In this work we have studied both Fe and N self-diffusion and
relaxation using SIMS and Fe diffusion using NR technique.
Simultaneous measurement of Fe and N self-diffusion in FeN are not
reported previously to the best of our knowledge. However such
measurements in a different class of material (Si$_{3}$N$_{4}$)
exist.~\cite{Schmidt:PRL:2006,Schmidt:AM:2008} While SIMS is an
established technique for measuring self-diffusion, neutron
reflectometry is an emerging technique and is of special interests
to measure small diffusion lengths. The depth resolution available
with SIMS is of the order of 5\,nm where as NR gives a depth
resolution as low as 0.1\,nm. Therefore SIMS is best suited when
involved diffusion lengths are large while information about small
diffusion lengths could only be obtained using NR. In this work we
have used both techniques in a complementary way.  Stable isotopes
of Fe and N namely $^{57}$Fe and $^{15}$N provide a very good
contrast as compared to their natural abundances with SIMS
technique. Also, Fe and $^{57}$Fe have very good contrast for
neutrons. The neutron coherent scattering length (SL) for natural
Fe is 9.45\,fm while for $^{57}$Fe is 2.3\,fm. The neutron SL for
natural N and $^{15}$N is 9.36\,fm and 6.44\,fm, respectively.
Given the low concentration of nitrogen in nitrogen poor phases
($\sim$14\%) and low contrast between N and $^{15}$N for neutrons,
N self-diffusion measurements with NR could not be performed.

\section{\label{sec:level2}Experimental}

Iron nitride thin films were prepared using a dc-magnetron
sputtering technique. A trilayer sample with nominal structure of
Si (100) /FeN(120\,nm)/$^{57}$Fe$^{15}$N(9\,nm)/FeN(120\,nm) was
deposited at room temperature (without intentional heating) for
SIMS measurements. The samples were deposited using a mixture of
Ar and N as sputtering gases with a ratio of 9:1, respectively.
The total gas flow was kept at 30\,cm$^{3}$/min. To deposit the
isotope marker layer of $^{57}$Fe$^{15}$N the $^{57}$Fe foil was
sputtered with a mixture of Ar and $^{15}$N gases while for
depositing FeN layer, natural iron target was sputtered with a
mixture of Ar and natural nitrogen gases. In order to minimize
intermixing of natural and $^{15}$N gases, a residual gas analyzer
(RGA) was installed in the sputtering system. After the deposition
with natural nitrogen gas, the system was pumped down while
monitoring the presence of the residual gases. When the pressure
of the natural nitrogen reached to a minimum in RGA, $^{15}$N gas
was allowed to flow in the chamber. Similarly for the next layer
with natural nitrogen, $^{15}$N gas was allowed to pump down. The
deposition of the film was carried out after obtaining a base
pressure better than 1$\times$10$^{-6}$ mbar. During the
deposition the pressure in the chamber was
5$\times$10$^{-3}$\,mbar. The sample was deposited at a constant
sputtering power of 50\,W. Before deposition the vacuum chamber
was repeatedly flushed with Ar gas so as to minimize the
contamination of the remaining gases present in the chamber. The
targets were pre-sputtered for about 5\,minutes in order to remove
possible surface contaminations. Similarly, Si(100)[FeN
(22\,nm)/$^{57}$FeN (8\,nm)]$_{10}$ multilayers were deposited for
NR measurements. In this case natural N and $^{15}$N gases and
natural Fe and $^{57}$Fe targets were alternatively sputtered to
prepare the multilayer with 10\,periods. All other deposition
parameters were kept identical as they were in case of the
trilayer sample.

The structural characterizations of the samples were carried out
with x-ray reflectivity (XRR) and grazing incidence diffraction
(GIXRD) using a standard x-ray diffractometer with Cu\,K-$\alpha$
x-rays. The microstructure of the films were obtained using
transmission electron microscopy (TEM) in the planar mode. For
this purpose a thin film of about 70\,nm thickness was specially
coated directly on to the carbon coated TEM grid. The deposition
conditions for this sample were kept similar as described earlier
in this section. The TEM measurements were performed at room
temperature as well as at high temperatures using an
\textit{insitu} heater inside the sample space of the TEM. The
composition of the film is obtained from x-ray photoemission
spectroscopy (XPS). The XPS spectrum was collected using Al
K-$\alpha$ radiation of 1486.6\,eV. The base pressure in the
chamber during the measurement was better than
2$\times$10$^{-9}$\,mbar. Before recording the XPS patterns, the
sample was sputtered with 4\,KeV Ar$^{+}$ ions for 120\,minutes
with a very small ion current of 10-15\,$\mu$A in order to remove
the surface contamination without changing the composition of the
sample. The annealing of the samples for diffusion measurements
were performed in a vacuum furnace with a base vacuum better than
1$\times$10$^{-6}$\,mbar. The temperature in the furnace was
controlled with an accuracy of 1\,K.

For diffusion measurements the concentration depth profile was
measured by a secondary ion mass spectrometer (SIMS) CAMECA-IMS5F.
The Primary ions used for sputtering were Cs$^{+}$ ions of energy
4\,KeV and the ion current was about 30\,nA. The secondary ions
were detected by a double focusing magnetic mass spectrometer.
Neutron reflectivity measurements were performed at AMOR
reflectometer at the Swiss spallation neutron source (SINQ), at
Paul Scherrer Institute, Switzerland.~\cite{Gupta_PJP04} The
reflectivity pattern was measured using different angular settings
in the time-of-flight mode.

\section{\label{sec:level3}Results}

\subsection{\label{sec:level3.1} Structures and Composition}

The samples under investigation in this study are a trilayer (TL)
and a multilayer (ML) sample. The nominal composition of these are
as follows:

TL: Si(substrate)/FeN (90\,nm)/$^{57}$Fe$^{15}$N (9\,nm)/FeN
(90\,nm)

ML: Si(substrate)/[FeN (22\,nm)/$^{57}$FeN (8\,nm)]$_{10}$

\begin{figure} \center
\vspace{-0.2cm}
\includegraphics [width=75mm,height=65mm]{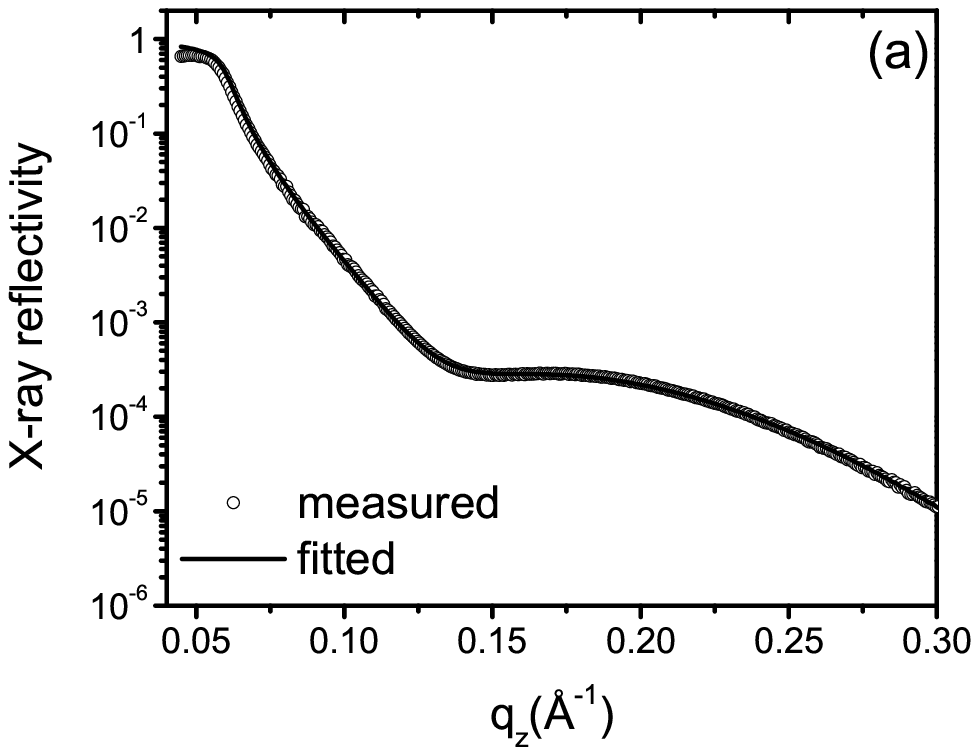}
\includegraphics [width=75mm,height=65mm]{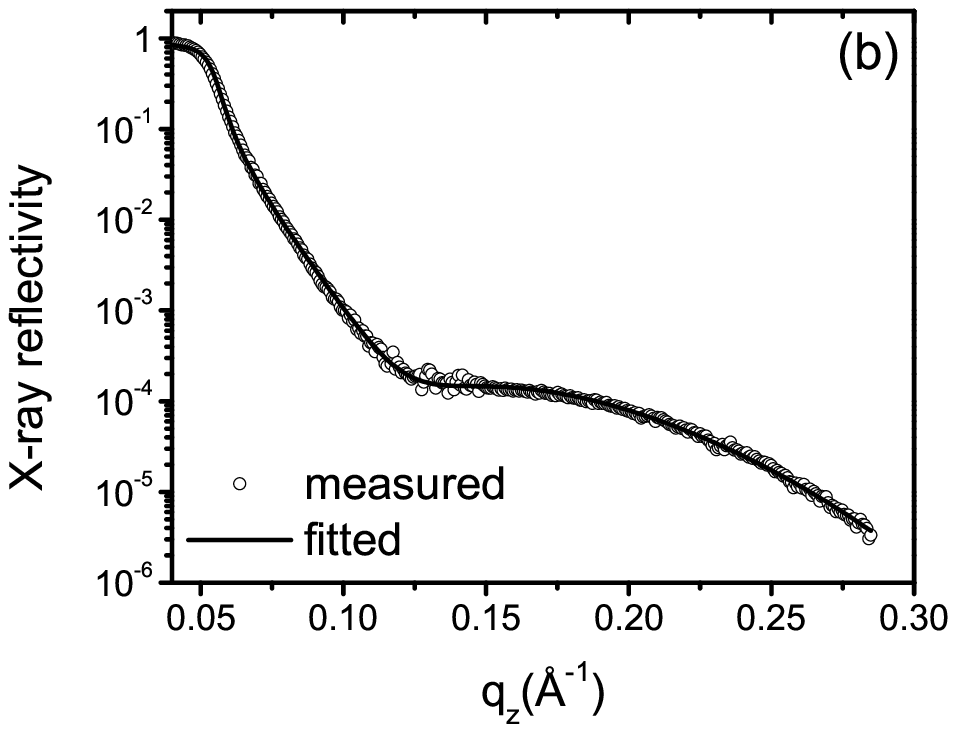}
\vspace {-0.2cm} \caption{\label{fig:fig1} X-ray Reflectivity
Pattern of the Si (substrate)/FeN (91.5nm.)/$^{57}$Fe$^{15}$N (7
nm)/FeN (91.5 nm) film (a) and Si(substrate)/[FeN
(22\,nm)/$^{57}$FeN (8\,nm)]$_{10}$ (b).}
\end{figure}

\begin{figure} [!hb]\center
\vspace{-0.2cm}
\includegraphics [width=75mm,height=90mm]{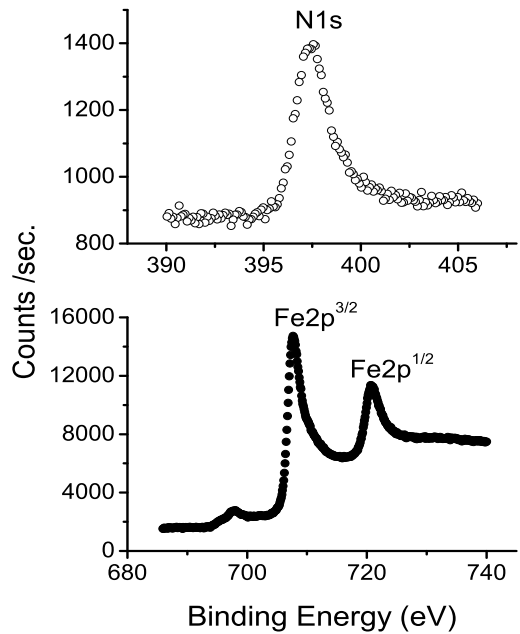}
\vspace {-0.2cm} \caption{\label{fig:fig2} XPS spectra of Si
(substrate)/FeN (91.5nm.)/$^{57}$Fe$^{15}$N (7 nm)/FeN (91.5 nm)
film. The binding energy of N1s (upper) and Fe2p$^{3/2}$,
Fe2p$^{1/2}$ (lower) are shown.}
\end{figure}

In both cases the chemical composition of FeN is the same and both
samples were prepared under identical condition of sputtering as
described in the previous section. In both samples, first x-ray
reflectivity (XRR) measurements were performed to cross check the
chemical composition. In case the composition of natural and
isotope layers is not the same, it would result in some structure
in XRR pattern due to a difference in scattering contrast.
Fig.~\ref{fig:fig1} shows the XRR pattern samples. As can be seen
from the figure the decay of the XRR intensity is `smooth' and
does not show any dominant feature. A diffuse maxima centered
around 0.17\,{\AA}$^{-1}$, observed in the pattern is due to the
formation of a oxide layer (about 3\,nm) at the surface of the
film. The feature-less pattern analogous to a single layer
confirms the chemical homogeneity of the samples and rules out any
variation in electron density being caused while switching to
$^{57}$Fe target from natural Fe and co-sputtering with mixture of
$^{15}$N and Ar gases during deposition of the isotopic marker
layer. Since the thickness of the sample being large the total
thickness oscillations could not be resolved due to limited
instrument resolution of the x-ray reflectometer. Total thickness
of the film (for TL sample) has been determined with the known
sputtering rate (calibrated for a thin film using XRR) and
cross-checked by measuring the depth of the crater formed after
SIMS measurements with DEKTEK depth profilometer. The total
thickness of the TL sample was found to be 190\,nm. Accordingly,
the actual structure of the film is Si (substrate)/FeN
(91.5\,nm)/$^{57}$Fe$^{15}$N (7\,nm)/FeN (91.5\,nm).

The chemical composition of the samples was obtained using XPS.
The binding energy of Fe2p$^{3/2}$, Fe2p$^{1/2}$ and N1s core
level corresponds to that of FeN as shown in Fig.~\ref{fig:fig2}.
The stoichiometry of the film as obtained from the XPS data is
Fe$_{86}$N$_{14}$. A small amount of oxygen is also detected which
was also seen from XRR measurements at the surface of the films.
Presence of this oxygen is due to absorbed oxygen from the
atmosphere due to exposure of the samples to the environment.

\begin{figure} \center
\vspace{-0.2cm}
\includegraphics [width=80mm,height=70mm]{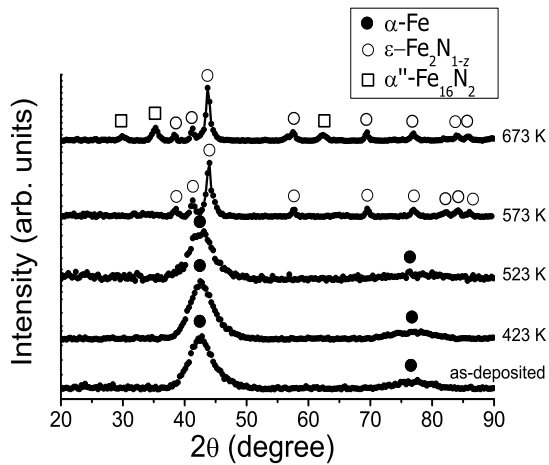}
\vspace {-0.5cm} \caption{\label{fig:fig3} Grazing incidence XRD
pattern of Si (substrate)/FeN (91.5\,nm.)/$^{57}$Fe$^{15}$N
(7\,nm)/FeN (91.5\,nm) film in the as-deposited state and after
annealing at different temperatures for 1 hour.}
\end{figure}

The structure of the sample was determined using XRD in the
as-deposited state and after annealing the sample at different
temperature. The XRD pattern of the as-deposited film as well as
the film annealed isochronally at different temperatures for
1\,hour is shown in Fig.~\ref{fig:fig3}. The broad diffuse maxima
at 2$\theta$=44$^{\circ}$, characteristic of an amorphous
$\alpha$-Fe phase is clearly visible. This corresponds to an
average Fe-Fe nearest neighbor distance $a=1.23\lambda
/2sin\theta$=2.5{\AA},~\cite{Guinier_XRD} where $\lambda$ is the
wavelength of x-rays (1.54 {\AA}) and $\theta$ is the mean
position where the diffuse maxima occurs. The obtained value
corresponds to that generally observed in transition
metal-metalloid (TM-$M$) glasses having a composition around the
eutectic phase. The crystallization behavior of the film was
studied by isochronal annealing of the film at different
temperatures.~\cite{Gupta:amorph_PRB05} From the figure it can be
seen that there is no appreciable change in the XRD pattern up to
a temperature of 523\,K. This indicates that the amorphous phase
is stable at least up to 523\,K. Annealing the film at 573\,K for
1\,hour causes crystallization and the phase formed is identified
as $\epsilon$-Fe$_{2}$N. Further annealing at 623\,K for 1\,hour
causes the formation of mixed phases, which is identified as
$\epsilon$-Fe$_{2}$N and $\alpha
^{\prime\prime}$-Fe$_{16}$N$_{2}$. Thus from XRD it can be
concluded that the system remain in amorphous state after
annealing up to a temperature of 523\,K. Therefore for
self-diffusion measurements we choose 498\,K as an upper limit of
annealing temperature. It was found that even after the highest
diffusion annealing temperature (i.e. 498\,K) and time the film
remained in the amorphous state.

\begin{figure*} \center
\includegraphics [width=170mm,height=120mm]{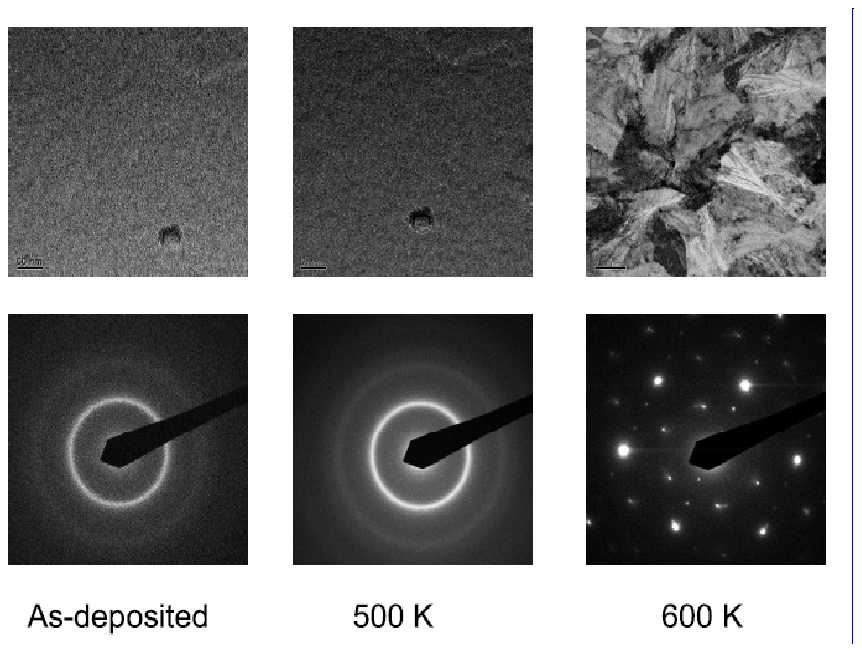}
\vspace{-0.2cm} \caption{\label{fig:fig4} TEM micrograph and
diffraction pattern of a FeN thin film of 70\,nm thickness
deposited on a C-coated TEM grid. The measurements were performed
in the as-deposited state and after heating the sample
\textit{insitu} at indicated temperatures. The bar size displayed
in the images is 50\,nm.}
\end{figure*}

\subsection{\label{sec:level3.2} Microstructure of FeN films}
The microstructure of the FeN film was investigated using TEM
measurements and shown in Fig.~\ref{fig:fig4}. As mentioned
earlier, for the TEM measurements a thin film of 70\,nm thickness
was directly deposited on a carbon coated TEM grid. The deposition
conditions for these films were kept identical to TL or ML
samples. In addition to TEM grid a sample was also deposited on a
Si wafer kept besides the grid. The thickness of the film coated
on Si substrate was measured using XRR and found to be 70\,nm. XRD
measurements of this film were also carried out and found similar
to that of TL or ML samples. It was found that the TEM micrograph
did not show the presence of grain or grain like structures and
the diffraction pattern showed diffuse rings indicating an
amorphous nature of the films. The samples were heated using a
heater and images were recorded \textit{insitu} at different
temperatures. Up to a temperature of 498\,K, there was no
appreciable change in the micrograph and diffraction pattern.
However after heating at 600\,K, the grains due to crystallization
of the amorphous phase appear and diffraction pattern showing
spots due to crystalline structure can be seen. The TEM
measurements were found in good agreement with XRD results
described earlier.

\subsection{\label{sec:level3.3} Diffusion measurements with SIMS}

With the structural, composition and microstructural
characterization of FeN films available, the thermal stability of
the films is known. Above a temperature of 523\,K, crystallization
of the films starts. Therefore for diffusion measurements a
temperature of 498\,K was chosen as a maximum temperature so that
diffusivity could be measured in the amorphous phase. In this
section results obtained on TL samples (Si /FeN
(91.5\,nm)/$^{57}$Fe$^{15}$N (7\,nm)/FeN (91.5\,nm)) using SIMS
are presented. Since both $^{57}$Fe and $^{15}$N are present as
the marker layer, depth profiling of $^{57}$Fe and $^{15}$N is
expected to peak at a position where marker layer is sandwiched.

\begin{figure} \center
\includegraphics [width=80mm,height=65mm]{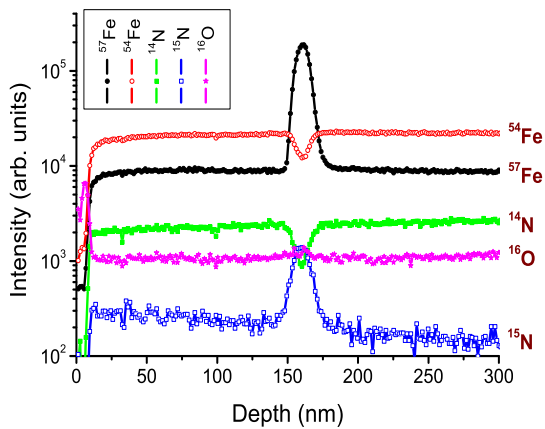}
\vspace{-0.2cm} \caption{\label{fig:fig5} (Color online) Secondary
ion mass spectroscopy (SIMS) depth profiles of the as-deposited
film.}
\end{figure}

Fig.~\ref{fig:fig5} shows SIMS depth profiles of $^{57}$Fe,
$^{54}$Fe, $^{15}$N, $^{14}$N and $^{16}$O in the as-deposited
film. At the position of the marker layer both $^{57}$Fe and
$^{15}$N show a peak while the concentration of $^{54}$Fe and
$^{14}$N shows a dip. The depth profile of $^{16}$O remains at a
constant level throughout the depth of the film. A closer look at
the depth profile of $^{57}$Fe and $^{15}$N reveals that the peaks
are some what skewed towards higher sputtering time or larger
depths. Such an asymmetry in the depth profiles is due to
radiation damage and small intermixing induced by the 4\,keV
Cs$^{+}$ ions used for sputtering the samples. A correction for
this irradiation broadening of the profiles is applied to the
primary concentration profiles. The concentration profiles are
corrected to yield the true ones according to the following
equation\cite{Brebee:Acta_Metall80,Limoge:JNCS:2000}
\begin{equation}
c_{r}(x+h) = c_{a}(x)+h {\frac{dc_{a}(x)} {dx} }, \label{eq:1}
\end{equation}
Where $c_{a}(x)$ and $c_{r}(x)$ are the experimentally determined
and true profiles, respectively, and $h$ is a parameter that
represents the strength of intermixing due to Cs$^{+}$ ion
bombardment. The value of $h$ was determined by applying this
correction on the as-deposited samples with known concentration
profile. The same value of $h$ was used for correcting the depth
profiles of the samples annealed for different periods of time.

\begin{figure} [!t]\center
\includegraphics [width=80mm,height=60mm]{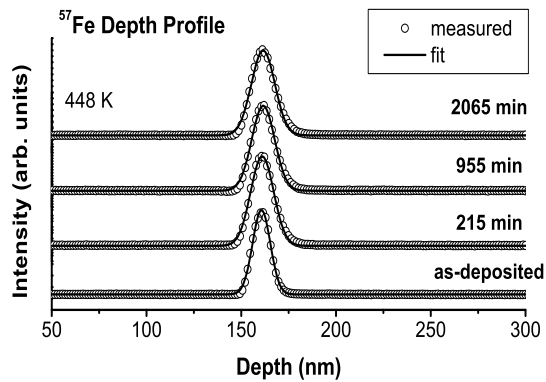}
\vspace{-0.2cm} \caption{\label{fig:fig6} SIMS depth profiles of
$^{57}$Fe in as-deposited sample as well as sample annealed at
448\,K for different times. The scattered points ($\circ$)
represents the experimental data and the solid line (--)
represents the fitted Gaussian profiles.}
\end{figure}

\begin{figure} \center
\includegraphics [width=80mm,height=72mm]{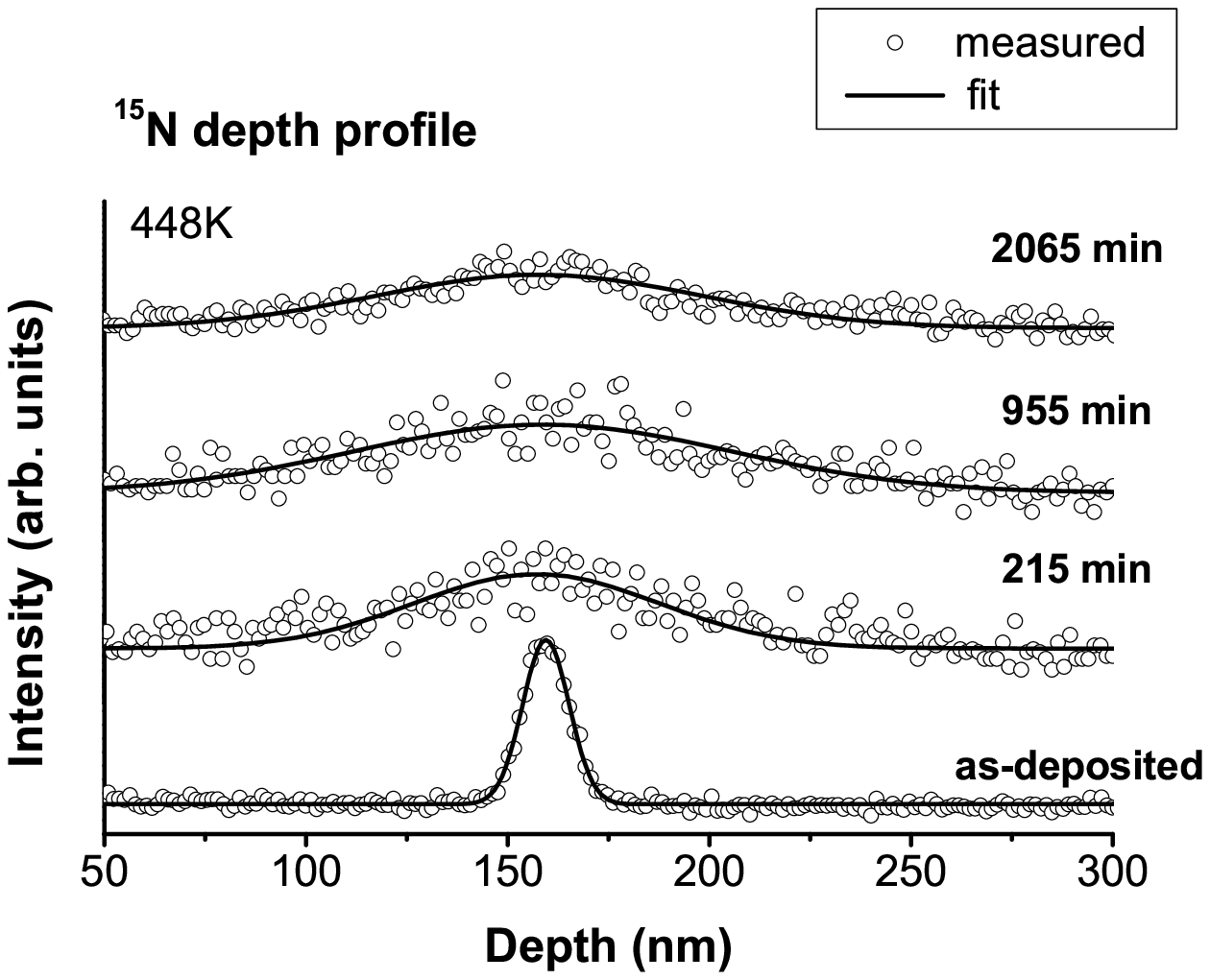}
\vspace{-0.2cm} \caption{\label{fig:fig7} SIMS depth profiles of
$^{15}$N in as-deposited sample as well as sample annealed at 448K
for different times. The scattered points ($\circ$) represents the
experimental data and the solid line (--) represents the fitted
Gaussian profiles.}
\end{figure}

For studying the self-diffusion of iron and nitrogen from the
diffusion mediated broadening of the depth profiles of $^{57}$Fe
and $^{15}$N respectively, isothermal annealing of the film was
performed at temperatures 448\,K and 498\,K for different times. A
typical broadening of the depth profile of $^{57}$Fe and $^{15}$N
as a function of annealing time at 448\,K is shown in
Fig.~\ref{fig:fig6} and Fig.~\ref{fig:fig7}, respectively. The
profiles have already been corrected for the Cs$^{+}$ ion
irradiation broadening. It may be observed clearly from
Fig.~\ref{fig:fig6} and Fig.~\ref{fig:fig7} that the diffusion
mediated broadening in the $^{15}$N profile is several times
larger as compared to the $^{57}$Fe profile, which indicates that
in this system nitrogen diffuses several times faster than iron.

\begin{figure*} \center
\includegraphics [width=85mm,height=80mm]{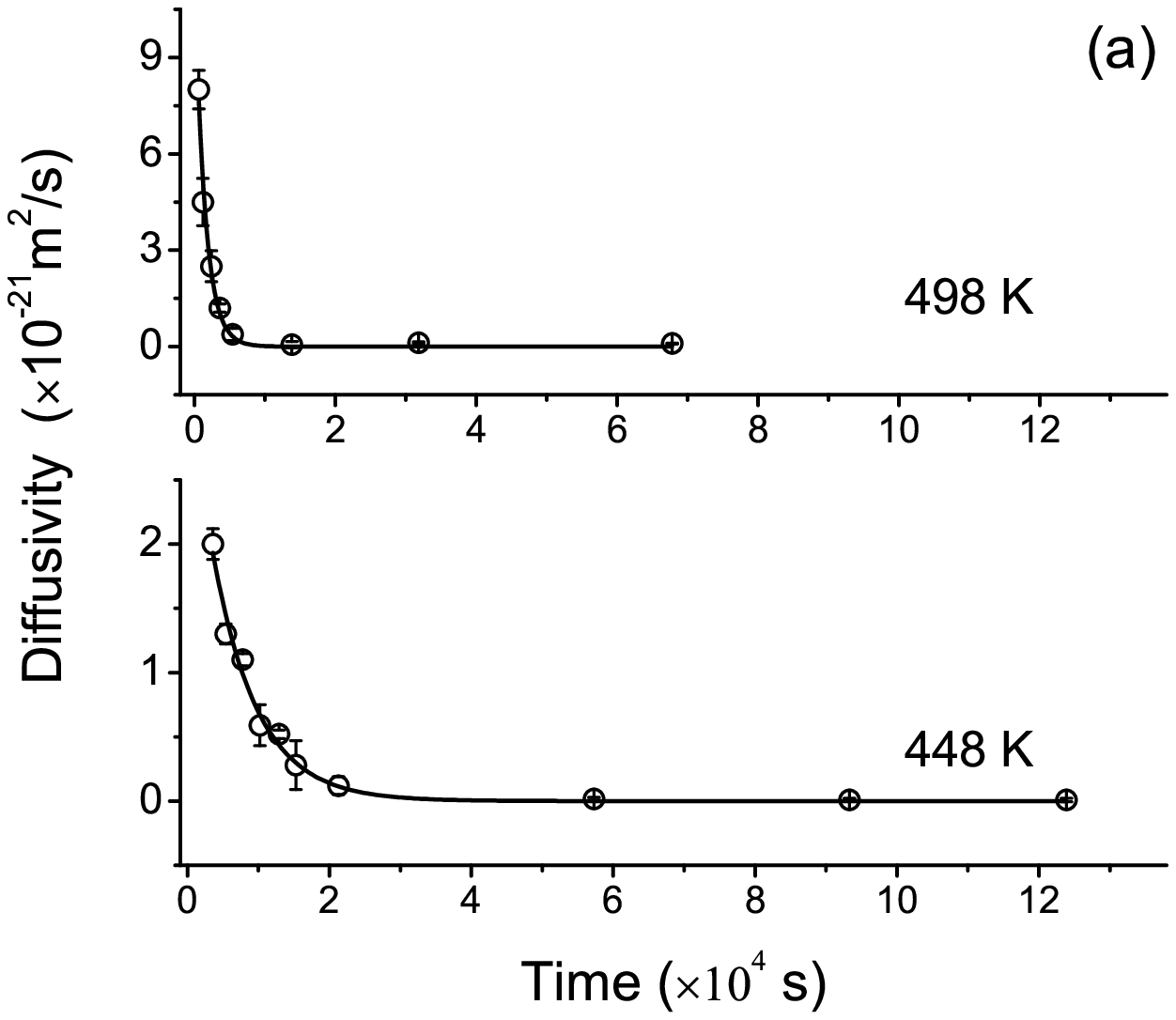}
\includegraphics [width=85mm,height=80mm]{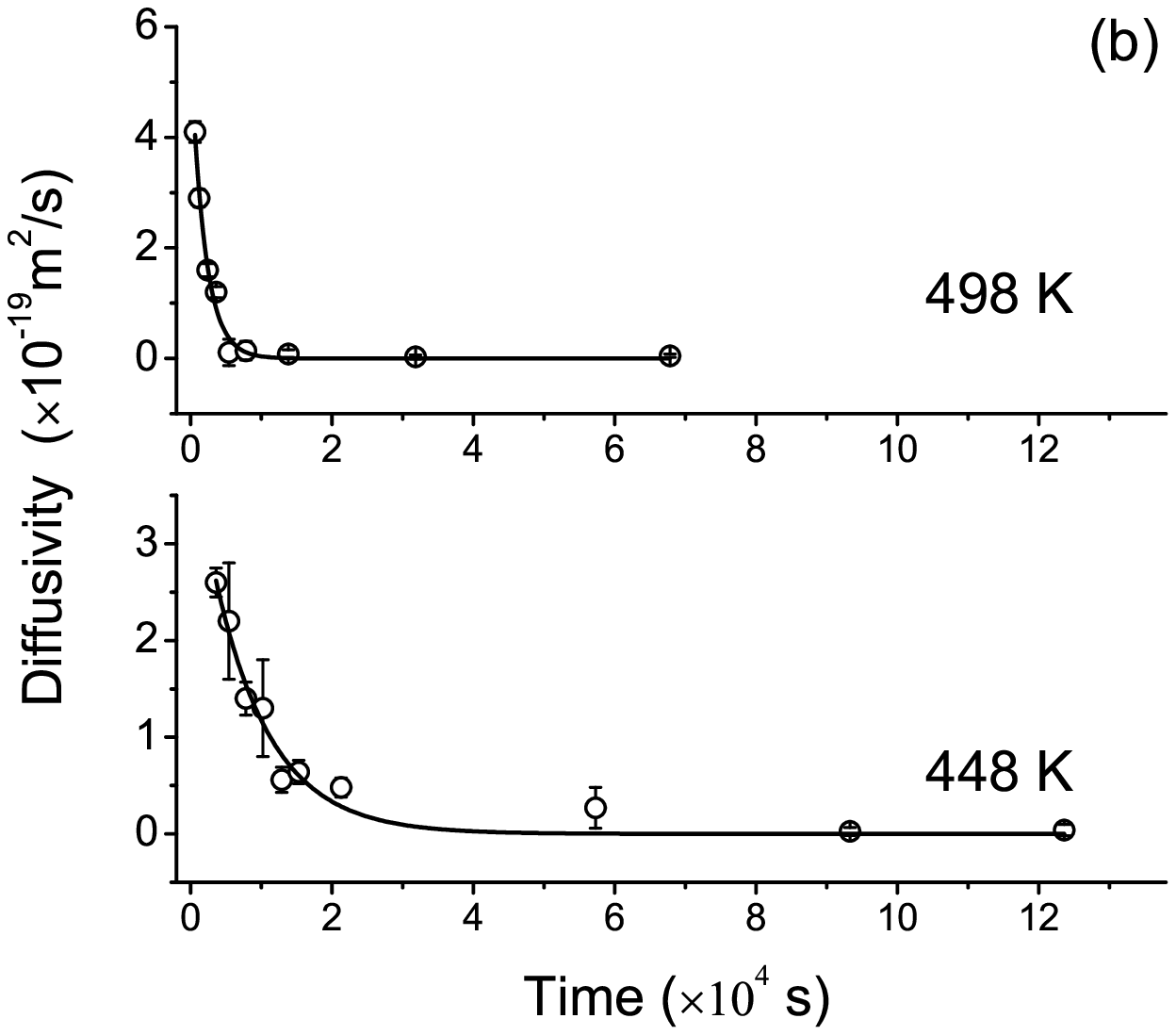}
\caption{\label{fig:fig8} Structural relaxation effects on the
time averaged diffusivity of (a) $^{57}$Fe and (b) $^{15}$N in FeN
at temperatures 448\,K and 498\,K. The error bars are typically of
the size of scattered data.}
\end{figure*}

In the present case, the thin film solution to Fick's law can be
applied and the tracer concentration $c(x,t)$ of $^{57}$Fe or
$^{15}$N as a function of penetration depth $x$ is given
by\cite{Shewmon}

\begin{equation}
c(x,t)=\textrm{const}  ~\textrm{exp}\Big( \frac {-x^{2}} {4Dt}
\Big), \label{eq:2}
\end{equation}

where $t$ is the time for annealing and $D$ is the diffusion
coefficient. Accordingly, the profiles were fitted to a Gaussian
and the time averaged diffusion coefficients were calculated using
the following equation~\cite{Limoge:JNCS:2000}

\begin{equation}
\langle D \rangle = \frac{\sigma^{2}_{t} - \sigma^{2}_{0}} {2t} ,
\label{eq:3}
\end{equation}

 where $\langle D \rangle$ is time averaged diffusivity and $\sigma_{t}$ is the
 standard deviation of the gaussian depth profile obtained after an annealing time of
 t and $\sigma_{0}$ is the standard deviation of as-deposited
 sample $(t=0)$.

In order to study the time variation of the diffusion coefficient,
diffusion measurements were carried out in the unrelaxed samples
by varying the annealing time at 448\,K and 498\,K.
Fig.~\ref{fig:fig8} shows obtained diffusivity as a function of
annealing time at these temperatures. As could be followed from
the figure, at the initial annealing time, the diffusivity decays
rapidly and becomes almost constant towards higher annealing
times. Such an annealing time dependence of the diffusion
coefficient is well known and attributed to structural relaxation
in the film.~\cite{Limoge:JNCS:2000,Tyagi:Acta_Metall_Mater91}

The observed time dependence of the diffusivity was fitted using
an exponential law for the relaxation and is given
by~\cite{Limoge:JNCS:2000}

\begin{equation}
\langle D \rangle (t) = A~\textrm{exp}(-t/\tau) + D_{SR},
\label{eq:4}
\end{equation}

Here $D_{SR}$ is the diffusivity in the structurally relaxed state
and $A$ is a constant. $A+D_{SR}$ gives the diffusivity at the
initial time $(t = 0)$, and $\tau$ is the structural relaxation
time. The solid line as shown in Fig.~\ref{fig:fig8}, gives a fit
of the measured data, yielding the value of $A$, $\tau$  and
diffusivity in the structurally relaxed state $D_{SR}$. As could
be seen from the figure, Fe diffusivity is about 2\,orders of
magnitude slower as compared to that of N at both the measured
temperatures. The relaxation times $\tau$ for Fe and N
self-diffusion at 448\,K and 498\,K are given in
table~\ref{tab:table1}. As could be seen from the table the
relaxation time in the two cases are comparable although the value
of diffusivity are different by about 2\,orders of magnitude for
Fe and N self-diffusion.
\begin{table}
\caption{\label{tab:table1} The relaxation time ($\tau$) for Fe
and N self-diffusion in FeN at 448\,K and 498\,K.}
\begin{ruledtabular}
\begin{tabular}{ccc}
Temperature & $\tau_{Fe}$ & $\tau_{N}$ \\
(K) & (s) & (s) \\ \hline

448 & 6252$\pm$450 & 7964$\pm$880\\

498 & 1509$\pm$256 & 2013$\pm$156\\

\end{tabular}
\end{ruledtabular}
\end{table}

The value of $D_{SR}$ comes out to be zero within the experimental
errors. This indicates that the absolute values of diffusivity in
the relaxed state are below the sensitivity of the SIMS technique.
Due to this reason we used neutron reflectivity technique to
measure the diffusivity of Fe in the relaxed state. As mentioned
already neutron reflectivity is a technique much better applicable
to measure small diffusion lengths due to a much better depth
resolution.

\subsection{\label{sec:level3.4} Diffusion measurements with NR}

Fe self-diffusivity in amorphous FeN was also measured using
neutron reflectivity. Neutron reflectivity, in particular, is
useful at low temperatures where diffusion distances are small and
can not be measured using SIMS depth profiling due to the limited
depth resolution. The depth resolution available with NR is of the
order of 0.1\,nm while that with SIMS is in excess of
5\,nm.~\cite{gupta:184206} One may also note that the value of
diffusivity for iron and nitrogen self diffusion in structurally
relaxed state is zero for both temperatures, which essentially
shows that the technique of SIMS is not sensitive enough to
determine such low diffusivities.  This also points out the
importance of neutron reflectivity measurements which have a much
higher sensitivity.

Since the samples under investigation in this study is nitrogen
poor (Fe$_{86}$N$_{14}$) and due to poor contrast between natural
N and $^{15}$N neutron scattering length (9.36\,fm and 6.6\,fm) N
self-diffusion using NR could not be studied. However measurements
of Fe diffusivity at low temperature would give additional
information about diffusion mechanism in this system. As mentioned
already NR measurement were performed on the ML sample
(Si(substrate)/[FeN (22\,nm)/$^{57}$FeN (8\,nm)]$_{10}$).

\begin{figure} \center
\includegraphics [width=75mm,height=90mm]{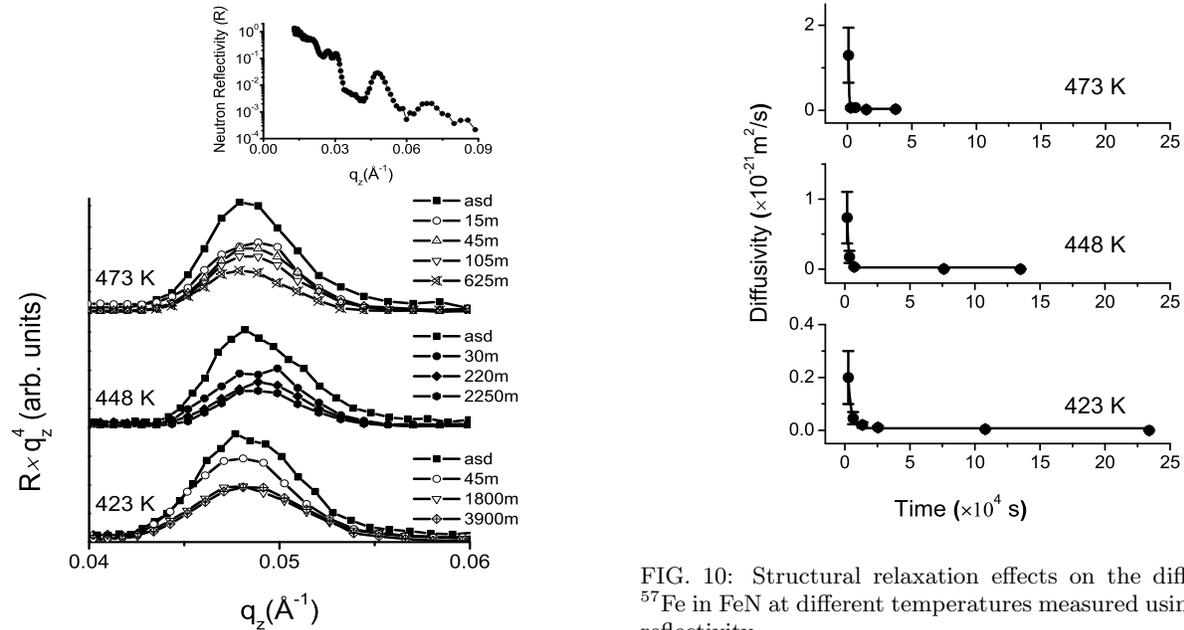}
\vspace{-0.5cm} \caption{\label{fig:fig9} Neutron reflectivity
measurements on Si(substrate)/[FeN (22\,nm)/$^{57}$FeN
(8\,nm)]$_{10}$ multilayer samples around second order Bragg peak
at different temperatures for different annealing times. The inset
of this figure shows the neutron reflectivity pattern of the
as-deposited sample.}
\end{figure}

NR measurements on the ML samples yield Bragg peaks due to
$^{57}$Fe periodicity. The inset of Fig.~\ref{fig:fig9} shows a
typical NR pattern obtained from such multilayers. It may be noted
that since FeN at this composition is magnetic, unpolarized NR
measurements yield critical edges due to reflection from domains
parallel to net magnetization of the samples and anti-parallel to
the direction of net magnetic moment. This effects results in two
critical edges and splitting of the Bragg peaks at lower q$_{z}$
values $(q_{z}= 4\pi/\lambda ~sin\theta)$, here $q_{z}$ is
momentum transfer and $\lambda$ is the wavelength of the used
radiation. However such splitting of the Bragg peak is reduced at
higher $q_{z}$ values. In the present study, second order Bragg
reflection occurring around $q_{z}$ = 0.048\,{\AA}$^{-1}$ was used
to measure Fe self-diffusion. The measurements were performed
using time of flight (ToF) mode and the angle of incidence
$\theta$ kept at 0.6$^{\circ}$ and 1.2$^{\circ}$ with wavelength
of incoming neutrons between 2\,{\AA}-10\,{\AA}.

\begin{figure} \center
\includegraphics [width=65mm,height=75mm] {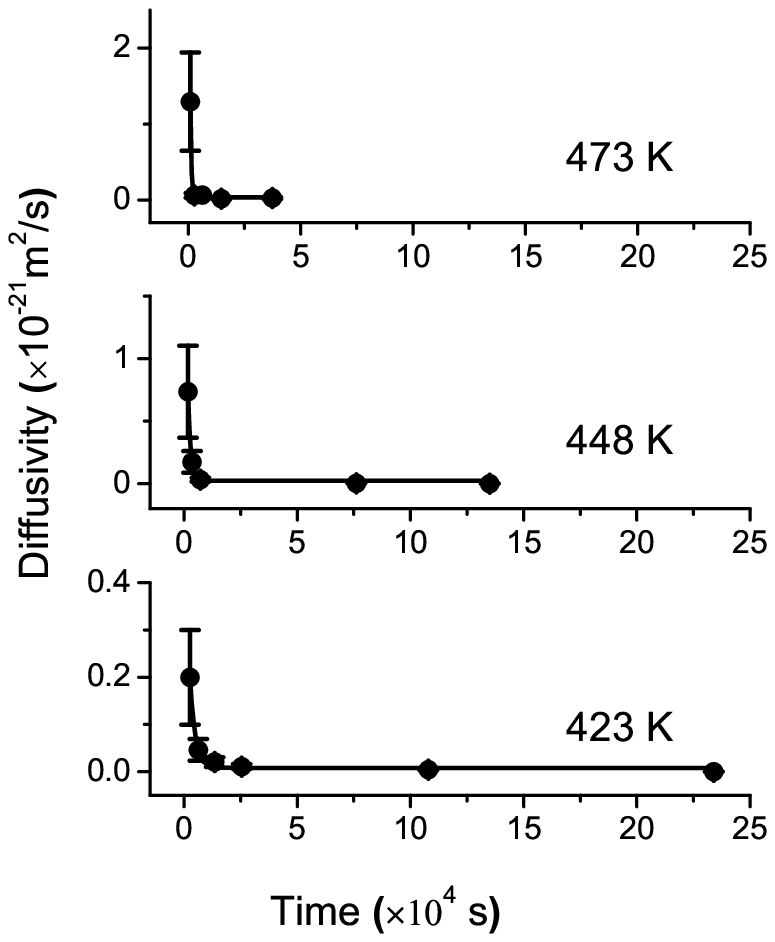}
\vspace{-0.2cm} \caption{\label{fig:fig10} Structural relaxation
effects on the diffusivity of $^{57}$Fe in FeN at different
temperatures measured using neutron reflectivity.}
\end{figure}

As the ML is annealed the intensity at the Bragg peak decays due
to diffusion. The decay of the Bragg peak intensity can be used to
calculate the diffusion coefficient using the
expression~\cite{Rosenblum_APL80,Speakman_JMMM96}

\begin{equation}
\label{eq:5} I(t) = I(0)\, \exp \left(-\frac{8 \pi^{2}\, n^{2}\,
D} {\ell^{2}}\,t\right),
\end{equation}

where $I(0)$ is the intensity before annealing and $I(t)$ is the
intensity after annealing time $t$ at temperature T, $\ell$ is the
bilayer periodicity $n$ is the order of reflection. The obtained
diffusivity using eq~\ref{eq:5} are plotted in
Fig.~\ref{fig:fig10}. Similar to results obtained using SIMS, Fe
diffusivity shows relaxation behavior. The diffusivity in the
structurally relaxed state are obtained using eq~\ref{eq:4}.

\section{\label{sec:level4}Discussions}

\begin{figure} \center
\includegraphics [width=75mm,height=75mm]{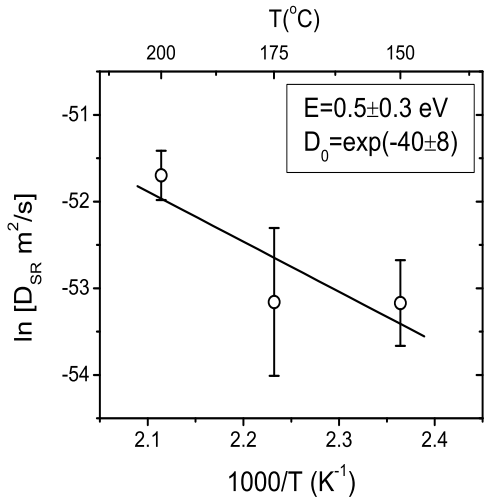}
\vspace{-0.5cm} \caption{\label{fig:fig11} (Color online)
Arrhenius behavior of Fe and N self-diffusion in FeN. Fe
diffusivity as obtained using SIMS and NR shows a good match
within the experimental errors.}
\end{figure}

\begin{figure} \center
\includegraphics [width=80mm,height=70mm]{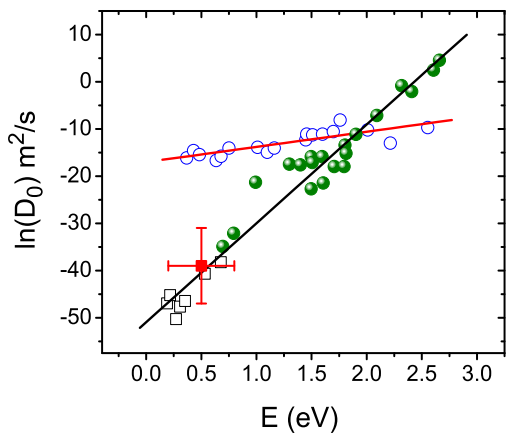}
\vspace{-0.1cm} \caption{\label{fig:fig12} (Color online) The
correlation between pre-exponential factor for diffusion D$_{0}$
and activation energy E. The literature values are taken from
reference.\cite{Wang:PRB:99,Utizg:JAP:1997,Hovarth,Sharma:AMM:1992,
Sharma:JMR:1989,Sharma:PRB:1992,Kronmueller:REDS:1989,Limoge:MSE:1997,Schober:PRB:1991}
The open symbol ($\circ$) corresponds to diffusion in crystalline
metals and alloys, the open square ($\square$) diffusion in
multilayers, the solid circle ($\bullet$) diffusion in
multilayers, the solid square ($\blacksquare$) Fe diffusion in
amorphous FeN (present study) and the solid line represents a
linear fit.}
\end{figure}

The results of Fe diffusion using SIMS and NR show relaxation
behavior. Also, N diffusion measured using SIMS show structural
relaxation. A plot of Fe diffusivity in the relaxed state obtained
using neutron reflectivity as a function of inverse of temperature
is shown in Fig.~\ref{fig:fig11}. A straight line fitting of the
experimental data using the expression $lnD  = lnD_{0} -E/k_{B}T$,
where $D_{0}$ is pre-exponential factor, $E$ is activation energy,
and $k_{B}$ is Boltzmann's constant, yields the values of $D_{0}$
and $E$, which are 0.5$\pm$0.3\,eV and exp(-40$\pm$8). It would be
interesting to compare the values of $D_{0}$ and $E$ obtained in
the present case with that for conventional amorphous alloys. It
is known that a correlation exits between $lnD_{0}$ and $E$, which
is universally followed in crystalline as well as amorphous
alloys, signifying the involved diffusion
mechanism.~\cite{Faupel_RMP03} Putting together the values
obtained in literature for bulk and conventional amorphous alloys
with the values obtained in the present case, it could be seen
from the Fig.~\ref{fig:fig12} that values obtained in the present
case goes well with the line obtained from literature data. The
relationship between $D_{0}$ and $E$ is known as isokinetic
relation and is given by~\cite{Linert:Chem_Soc_Rev:1989}

\begin{equation}
D_{0} = A ~\textrm{exp}(E/B) , \label{eq:6}
\end{equation}

where $A$ and $B$ are constants. As mentioned, this relationship
was found to be valid not only in amorphous alloys but also for
self and impurity diffusion in crystalline alloys involving both
substitutional and interstitial solid solutions. Though the values
of fitting constants $A$ and $B$ were found to be very different
in amorphous and crystalline alloys. In case of amorphous alloys
typical values of $A$ and $B$ are 10$^{-20}$ m$^{2}$/s and 0.055
eV, respectively while in case of crystalline alloys these are
10$^{-7}$\,m$^{2}$/s and 0.41\,eV, respectively. Putting together
the values of D$_{0}$ and E in the present case, the values of
constant $A$ and $B$ in the present case are
1.4$\times$10$^{-22}$\,m$^{2}$/s and 0.05\,eV, respectively, which
are very close to the values obtained for bulk and conventional
amorphous alloys. This is very interesting and it suggests that
even though the present system has much stronger covalent bonds
between iron and nitrogen nevertheless the diffusion mechanism is
very similar to that in other metallic glasses having metallic
bonds. In other words, in the present system the diffusion
mechanism is expected to be collective in nature involving many
atoms diffusing together. It is known from Zener's theory of the
pre-exponential factor that the pre-exponential factor could be
expressed as\cite{Shewmon,Faupel_RMP03}

\begin{equation}
D_{0}=ga^{2}f\nu_{0} ~\textrm{exp}(\Delta S/k_{B}), \label{eq:7}
\end{equation}

where $g$ is a geometric factor, $a$ the effective jump distance,
$\nu_{0}$ the effective jump attempt frequency, $f$ the
correlation factor, and $\Delta S$ the entropy of diffusion. Using
eq~\ref{eq:6}, the values of constant $A$ and $B$ could be written
as

\begin{equation}
A=ga^{2}f\nu_{0} ~\textrm{and}~ B = \frac{k_{B}E} {\Delta S},
\label{eq:8}
\end{equation}

With the known values of $B$ and $E$, the values of entropy,
$\Delta S$ for Fe diffusion is 10$k_{B}$. The values of $\Delta S$
in amorphous alloys ranges from 19$k_{B}$ to 56$k_{B}$ while in
crystalline alloys 2.5$k_{B}$ to 7.5$k_{B}$~\cite{Faupel_RMP03}.
In case of inter-diffusion in multilayers the values of $\Delta S$
are found in the range of 8$k_{B}$ to
15$k_{B}$~\cite{Wang:PRB:99,gupta:184206} which is very close to
the value found for Fe diffusion in the present case.

Concerning N diffusion, it is very clear that N diffusion is
faster compared to Fe diffusion. At this composition N occupies
interstitial sites in the crystal structure of
Fe.~\cite{Gupta:amorph_PRB05} It is well known that in the dense
random packing model for the metallic glasses, the metal atoms
form a dense random packing structure and the metalloid atoms
occupy the interstitial sites in these
structures.~\cite{Polk_Acta_Metal:1972} It has been reported that
TM-M systems have a high stability around a metalloid composition
of $\sim$20\%, as around this composition most of the interstitial
sites are occupied by metalloid atoms.~\cite{Polk_Acta_Metal:1972}
In the present case the metalloid composition (i.e. nitrogen) is
less than 20\% and therefore lot of interstitial sites remains
vacant. Thus, a very large number of vacant interstitial sites is
available in the system, which acts as an easy path for metalloid
atoms (i.e. nitrogen) to diffuse in the system. This might be the
possible reason for fast nitrogen self-diffusion. On the other
hand, the iron population in the system is very high and they are
very densely packed and there is no easy way for iron to diffuse
and also there is very least possibility of vacant space for iron
to diffuse. In other words the potential barrier around iron atoms
is very high and to overcome this barrier it needs a very high
energy. This might be to be the reason for the lower diffusion of
Fe as compared to N in the present system.

Looking further at Fig. 8, it is interesting to note that although
the diffusivity of Fe and N defer by orders of magnitude, the
relaxation time in the two cases are comparable.  It may be noted
that diffusivity in the amorphous materials decreases with
annealing time essentially because of structural relaxation.With
thermal annealing the amorphous structure evolves towards the more
stable state with lower free energy. This involves annihilation of
quenched-in excess free volume as well as topological relaxation
of the structure.~\cite{Limoge:JNCS:2000,Franck:DDF:1997} This
structural relaxation would affect the diffusivity of both Fe and
N atoms. Thus the relaxation time obtained from the diffusivity
measurements essentially pertains to the structural relaxation of
the amorphous phase.  And therefore, relaxation times as obtained
from Fe diffusion as well as N diffusion should be comparable.

\section{Conclusion}
In conclusion, the self-diffusion of Fe and N in an amorphous
Fe$_{85}$N$_{15}$ alloy has been studied. The Fe diffusion over a
wide temperature range has been studied by combining neutron
reflectivity and SIMS techniques while N diffusion was measured
using SIMS.  In case of Fe diffusion the observed relations
between activation energy and the pre-exponent factor suggest a
diffusion mechanism involving a large group of atoms, similar to
other metal-metalloid and metal-metal alloys. N diffusivity is
found to be several orders of magnitude higher as compared to that
of Fe.  This may be attributed to a diffusion mechanism for N
atoms involving jumps through the vacant interstitial sites in the
Fe network. Despite a large difference in the values of Fe and N
diffusivities, the relaxation time for the decay of diffusivity
with annealing time comes out to be same in the two cases,
suggesting that the structural relaxation in the amorphous phase
effects from the diffusion of both Fe and N in a similar manner.

\section*{ACKNOWLEDGEMENT}

A part of this work was performed at the Swiss Spallation Neutron
Source, Paul Scherrer Institute, Villigen, Switzerland. Partial
support from Indo-French Center for Promotion of Advanced Research
is acknowledged. Thanks are due to Y. L. Chiu for the help
provided in TEM measurements.

\bibliography {Amorphous_Fe_N_Diff}

\end{document}